\documentclass[12pt,preprint]{aastex}

\newcommand{\fwhm}{{\small FWHM}}

\newcommand{\chandra}{{\it Chandra}}

\newcommand{\hetgs}{{\small HETGS}}

\newcommand{\x}{X-ray}

\newcommand{\cmsq}{cm$^{-2}$}
\newcommand{\kms}{km~s$^{-1}$}
\newcommand{\etc}{$\eta$~Car}
\newcommand{\msol}{$M_{\odot}$}
\newcommand{\msolyr}{$M_{\odot}$~yr$^{-1}$}
\newcommand{\etal}{et~al.}

\begin{document}

\title{A hot transient outflow from $\eta$ Carinae}
\author{Ehud Behar\altaffilmark{1}, Raanan Nordon\altaffilmark{1},
Eyal Ben-Bassat\altaffilmark{1}, and Noam Soker\altaffilmark{1}}

\altaffiltext{1}{Department of Physics,
                 Technion, Haifa 32000, Israel.
                 behar@physics.technion.ac.il (EB).}

\shorttitle{Outflow from $\eta$ Carinae} \shortauthors{Behar et
al.}

\begin{abstract}
$\eta$~Carinae (\etc) is a stellar binary system with a period of
5.54 years.  It harbors one of the brightest and most massive
stars in our galaxy.  This paper presents spectroscopic evidence
for a fast (up to 2,000 \kms) X-ray outflow of ionized gas
launched from \etc\ just before what is believed to be the binary
periastron (point of smallest binary separation). The appearance
of this high-velocity component, just as the irregular flares in
the \x\ light curve, can not be explained by the simple continuous
binary wind interaction, adding to the intrigue of the \etc\
system.
\end{abstract}


\section{INTRODUCTION}
\label{sec:intro}

Despite being studied for many years and in all wavebands, the
$\eta$ Carinae system (\etc), best known for its spectacular
nebula \citep{morse98} that can be tracked back to the Great
Eruption of 1840 \citep{davidson97} is still much of a mystery. It
is now believed to be a stellar binary system with a period of
5.54 years \citep{damineli96}. The primary star is one of the
brightest and most massive ($\sim$ 120~\msol) stars in our galaxy.
The system is currently in a phase of immense mass ejection. The
\x s detected from \etc\ can be explained by the collision of the
two stellar winds \citep{corcoran01, pittard02}, not so the 70 day
X-ray minimum observed periodically around periastron
\citep{corcoran05}.

Currently, the mass loss rate of the primary exceeds
$10^{-4}$\msolyr\ \citep{smith03} and that of the secondary is an
order of magnitude lower \citep{pittard02}, but still unusually
high. The wind of the secondary is much faster \citep[few 1,000
\kms,][]{pittard02} than that of the primary \citep[$\sim$~600
\kms\ and depending on latitude,][]{smith03}. The \x\ flux between
2 and 10~keV during the past 12 years is normally at a level of
5$\times 10^{-11}$ erg~s$^{-1}$~cm$^{-2}$, as expected from the
collision of the two stellar winds \citep{usov92, corcoran01,
pittard02}. The \x\ light curve is roughly constant throughout
most of the 5.54 year orbital period, but rises gradually by a
factor of 3--4 upon approach to periastron at which time short
flares appear \citep{corcoran97}, before it drops sharply and
stays in an \x\ low state ($<$~20\% of normal brightness) for
approximately 70~days. The full X-ray light curve can be found in
\citet{corcoran05}. Although absorption by the dense primary wind
may play a certain role in attenuating the X-ray emission from
\etc, both spectral and temporal arguments have been put forward
to reject absorption or an eclipse as the prime reason for this
persisting low state \citep{hamaguchi04, soker05, akashi06}, which
is unique to \etc. Recently, it has been proposed that if the
binary separation during periastron is small enough, the massive
primary wind could smother the secondary wind and thus shut down
the X-ray emission \citep{soker05, akashi06}. However, so far
there has been no direct evidence for accretion.

In this work, we wish to add to the temporal aspects of the \etc\
tale a high-resolution \x\ spectroscopic dimension. This is
achieved by using five deep exposures of \etc\ with the High
Energy Grating Spectrometer (\hetgs) on board the \chandra\ \x\
Observatory (CXO), only the first of which has been published
\citep{corcoran01,corcoran01b}. Theoretical variations of spectral
line profiles in binary colliding wind systems have been modeled
by \citet{henley03}. The varying line profiles presented here for
\etc, however, do not follow these models, at least not in a
straightforward way, as we explain in \S 3.

\section{Observations \& Results}

The log of the \chandra\ observations is given in Table~1.   Two
observations (3745 and 3748) coincide with the short intense
flares just before periastron.  The exact time of periastron is
not well determined and depends on the uncertain orbital
parameters of the system. Nonetheless, for the sake of
presentation, we assume throughout this paper a period of
5.54~years (2024~days) and set phase zero to be at 2003-06-29. The
X-ray low state, thus, occurs during phase 0 -- 0.035. The data
were retrieved from the CXO archive and processed using standard
software (CIAO version 3.2.2.). \chandra 's HEG and MEG (high and
medium energy gratings) spectrometers operate simultaneously in
the 3.5 -- 7.5 \AA\ band of interest here and their spectra are
consistent with each other. Aiming at the highest possible
kinematic resolution, however, the higher spectral resolving power
of the HEG in this band (factor of $\sim$ 2) is favored over the
slightly higher effective area of the MEG (factor of $\sim$ 2),
especially since the co-added plus and minus 1st order HEG data
are already of sufficiently high signal-to-noise ratio (S/N).

To reveal the dynamics of the \x\ gas, we transferred each
spectral line profile to line-of-sight velocity space ($v_\|$ in
\kms) using the simple non-relativistic Doppler shift $v_\| =
c(\lambda - \lambda _0)/ \lambda _0$, where $\lambda$ and $\lambda
_0$ represent the observed and rest-frame wavelengths of the line,
respectively, and $c$ is the speed of light. Negative and positive
velocities represent approaching and receding gas, respectively.
Fig.~1 shows the broad and variable \x\ line profiles observed in
\etc\ as demonstrated on the Si$^{+13}$ Ly$\alpha$ unresolved
doublet at 6.18~\AA. As expected, the MEG profile represents a
smoothed version of the higher-resolution HEG profile. The
observed line is much broader than the instrumental line spread
function (also shown). Fig.~1 also demonstrates how the line
profile varies dramatically from assumed phase $\phi$=~--0.470
(Obs. 632) to $\phi$=~--0.028 (Obs. 3745). The early-phase
($\phi$=~--0.470) line is centered around zero velocity with no
significant emission beyond 700~\kms\ (\fwhm\ = 1000
$\pm$~100~\kms). The line closer to periastron ($\phi$=~--0.028)
features high velocity gas up to $\sim$2000~\kms. A gaussian fit
yields a centroid shift of --550 $\pm$~90~\kms, although the
profile is clearly asymmetric.


In each observation, the kinematic profiles of all lines are
fairly similar. This is demonstrated in Fig.~2, where we present
four bright spectral lines from Obs.~3745 ($\phi$= --0.028). The
profiles are unambiguously broad and asymmetric with approaching
velocities reaching 2,000~\kms, but with no significant receding
emission beyond $\sim$~700~\kms. In order to understand how the
line profiles vary between observations and as a function of phase
we co-added the profiles of nine bright lines in each observation,
namely the Ly$\alpha$ (1s-2p), He$\alpha$ resonance (1s$^2$-1s2p)
and He$\alpha$ forbidden (1s$^2$-1s2s) lines of the Si, S, and Ar
ions. Since the profiles of all these lines in a given observation
are roughly similar (e.g., Fig.~2), co-adding gives an average
profile with improved S/N. Individual Fe lines below 2~\AA\ could
not be included in this analysis, despite their high intensity,
since the \hetgs\ resolving power decreases strongly at these low
wavelengths to the extent that the He-like Fe line complex is
unresolved.  The blue wing of the He$\alpha$ Fe resonance line is
consistent with, though not resolved as well as, the other line
profiles presented here. The resulting mean profiles have been
normalized (except for Obs. 3747) to facilitate the comparison and
are presented in Fig.~3. The observed, mean, unnormalized peak
intensities for the assumed phases: --0.470, --0.130, --0.028,
--0.006, and +0.044 are, respectively, 2.0, 3.9, 3.9, 1.9, and
0.3, $\times$~10$^{-3}$~ph~s$^{-1}$~cm$^{-2}$~\AA$^{-1}$. However,
any interpretation of the {\it absolute} line fluxes in terms of
the orbital parameters would be risky, as the observed variability
is a result of both abrupt flaring and varying absorption.

A clear trend can be seen in Fig.~3. Far from periastron, the line
profile is relatively symmetrical and narrow. Closer to
periastron, bright components of gas moving towards us at
velocities as high as --2,000~\kms\ start to develop. As the
system further approaches periastron, the outflow dominates the
line profile to the point where the bulk of the emission is
clearly blue-shifted. After periastron absorption considerably
attenuates the emission lines to the low level seen in Fig.~3.
Naive gaussian fits to the first two profiles yield centroids at
--25 $\pm$~25~\kms\ (for $\phi$=~--0.470) and at --50
$\pm$~20~\kms\ ($\phi$=~--0.130) both with \fwhm\ = 700
$\pm$~50~\kms. The late-phase asymmetric profiles are
non-gaussian. For the mere purpose of quantifying the centroid
shift, we performed two-gaussian fits, which yield peaks at --120
$\pm$~50~\kms\ and --800 $\pm$~100~\kms\ for $\phi$=~--0.028 and
at --170 $\pm$~50~\kms\ and --1000 $\pm$~100~\kms\ for
$\phi$=~--0.006. It is hard to rule out high-velocity {\it
receding} gas before the \x\ minimum, since in agreement with
\citet{hamaguchi04}, we find that the column density toward the
\x\ source rises from 5$\times$10$^{22}$~\cmsq\ long before
periastron to 3$\times$10$^{23}$~\cmsq\ just after it. Emission
from the far side of the system, thus, could be strongly absorbed.


\section{Interpretation \& Discussion}

The colliding wind binary model, which is generally accepted for
explaining the \x\ emission of \etc, has several clear predictions
regarding the spectral line profiles expected at different orbital
phases.  The Doppler shifts are the result of shocked gas flowing
along the contact discontinuity (CD) surface and away from the
stagnation point (SP). If the shock opening angle is between
$45^\circ\ - 90^\circ$ \citep{pittard02}, the broadest line
profiles are expected when the system is observed perpendicular to
the line connecting the two stars, as the hottest gas formed
around the SP flows directly towards and away from the observer.
In this case, the lines are centered at zero velocity and are
intrinsically symmetrical, but could be skewed by absorption of
the far (red) emission.  On the other hand, when the system is
observed from behind one of the stars, shifted centroids are
expected. If the system is observed from behind the weaker
(stronger) wind, downstream shocked gas is flowing along the CD
surface in the general direction of (opposite) the observer,
producing blue- (red-) shifted centroids. The maximum velocities
in this case would be less than those in the first (orthogonal)
orientation. For a complete and detailed calculation of line
profiles from colliding winds see Henley et al. (2003). In the
following, however, we argue that the current line profiles
observed at the different phases of the \etc\ orbit are in
contradiction, even qualitatively, with the simple colliding wind
scenario.

If the major axis of the \etc\ binary is perpendicular to our line
of sight \citep{smith04}, the broadest profiles are expected at
apastron (see above), but that is when the observed profiles are
narrowest ($\phi =-0.470$).  Also, with this geometry, the
secondary would have to pass in front of the primary before
periastron (in contrary to the picture of \citet{smith04}) to
explain the blue-shifted centroids at $\phi = -0.028, -0.006$, but
that would produce symmetrical (blue-shifted) profiles as
absorption through the secondary wind is weak, while strongly
asymmetrical profiles are observed.  In any case, the lower
velocities observed near apastron appear to rule out this
orientation if the colliding wind geometry is to produce the
observed \x\ line profiles.

If alternatively, the projection of the line of sight on the
orbital plane is more or less along the major axis and the viewing
angle at apastron is from behind the secondary \citep{corcoran01},
the observed blue-shifts at $\phi = -0.028, -0.006$ could be
ascribed to gas flowing from the SP towards the observer and the
unobserved red wing of the line to absorption by the primary wind.
However, according to this scenario, at apastron one would expect
blue-shifted centroids of at least a few 100~\kms\ due to
downstream gas (see above), which is not observed ($\phi =
-0.470$). For an observer above the binary plane ($i < 90^\circ$),
as seems to be the case in \etc\ \citep{corcoran01}, the problem
becomes more severe (higher blue-shift is expected, but not
observed).

In short, the observed behavior of the \x\ line profiles of \etc\
appear qualitatively inconsistent with a naive wind-collision
scenario, since the interpretation of the high blue-shifts from a
post-shock stream can not be made consistent with all phases.
\citet{henleyphd}, comparing with line profiles calculated from
hydrodynamical simulations, also reached the conclusion that the
observed profiles are not as expected from simple models of the
geometry of the wind-wind collision. He suggested that including
the Coriolis effect may bring the models to better agreement with
observations. \citet{kashi07} suggest the effect of orbital
motion, which can account for small centroid shifts, but can not
explain the currently observed \x\ profiles. Here, we wish to
propose an alternative explanation. In particular, we note that
the asymmetric line profiles are observed during strong peaks in
the X-ray light curve. These peaks too can not be accounted for by
the simple colliding wind model \citep{pittard02, henley03}, or by
the Coriolis force.

We therefore prefer to interpret the observed high-velocity gas up
to --2000~\kms\ in terms of a transient, collimated fast wind, or
a jet, ejected from the immediate vicinity of the binary system
through the interaction of the two stars. The peaks in the light
curve hint that the outflow may be in the form of blobs. The
unresolved \chandra\ images and the rapid variations both indicate
that the outflow is restricted to within approximately
2$\times$10$^{16}$~cm from the center (assuming 0.5\arcsec\ at
\etc 's distance of 2.3~kpc). The appearance of the same charge
states in the spectra attained during all phases implies that the
temperatures of the X-ray gas and line-emitting outflow remain in
the range of $kT$~= 2~-- 5~keV. Consequently, we speculate that
the fast component of the outflow consists of gas shocked by the
collision of the winds (as observed throughout the orbit) that is
launched and collimated near periastron. The widths of the major
peaks seen in Figs.~2 and 3 suggest that the outflow is only
moderately collimated to within $\sim 30^\circ$. The condensation
of shocked gas provides a natural explanation for the rise in
X-ray intensity during the short intermittent flares observed in
the X-ray light curve. A collimated outflow from the secondary has
also been suggested to explain the enhanced He~II $\lambda 4686$
emission before periastron \citep{soker06}. Interestingly, both
proper motion and Doppler shifts have been measured for
high-velocity visible-light knots much further out from the center
\citep{walborn78, meaburn93}. The optical knots were ejected more
than a hundred years ago along the minor axis of the nebula
\citep{meaburn93}. We would expect the present outflow to be
launched along the major axis of the nebula although we have no
pertinent information on its present direction with respect to the
system's geometry. If indeed that is the case and if the
inclination angle of the binary plane to our line of sight is $i=
42^\circ$ \citep{smith02}, the actual outflow velocity would be as
high as $\sim$2,700~\kms.


As the massive \etc\ primary is a short-lived star shedding
considerable mass and as it is now understood that core collapse
supernovae can not expel more than a few solar masses, implying
that higher mass stars must shed most of their mass prior to the
explosion, the winds and fast outflow of \etc\ may be a supernova
in the making. At the very least, it highlights the next
periastron passage on 2009, January 12 as a faithfully scheduled
experiment for the astrophysically common, but poorly understood
phenomenon of jet launching.

\acknowledgments This research was supported by grant \#28/03 from
the Israel Science Foundation, and by a grant from the Asher Space
Research Institute at the Technion.

\clearpage

\begin{deluxetable}{lcccc}
\tablecolumns{6} \tablewidth{0pt} \tablecaption{CXO/HETG
Observation Log \label{tab1}} \tablehead{
   \colhead{CXO archive} &
   \colhead{Start Time} &
   \colhead{Exposure} &
   \colhead{Assumed Orbital} &
   \colhead{2--10~keV Flux} \\
   \colhead{ID \tablenotemark{a}} &
   \colhead{} &
   \colhead{(ks)} &
   \colhead{Phase ($\phi$)\tablenotemark{b}} &
   \colhead{(10$^{-10}$ erg~s$^{-1}$~cm$^{-2}$)}}
\startdata
632 & 2000-11-19 02:46:40 & 90.69 &  --0.470 & 0.50 \\
3749 &  2002-10-16 08:08:49 & 93.96 &  --0.130 & 0.98 \\
3745 &  2003-05-02 11:56:16 & 97.29 &  --0.028 & 2.2 \\
3748 &  2003-06-16 05:35:28 & 100.1 &  --0.006 & 0.97 \\
3747 &  2003-09-26 22:45:53 & 72.16 &  +0.044 & 0.48 \\
\enddata
\tablenotetext{a}{ One additional observation was carried out
during the X-ray minimum (ID 3746 at 2003-07-20 01:46:22) for
which no useful spectrum could be obtained.} \tablenotetext{b}{
Phase is approximated based on the assumption of a 5.54~year
(2024~day) orbit, where phase zero is set at 2003-06-29 and the
X-ray low state occurs during phase 0 -- 0.035.}
\end{deluxetable}

\clearpage

\begin{figure}
  \centering
  \includegraphics[width=10.0cm, angle=0]{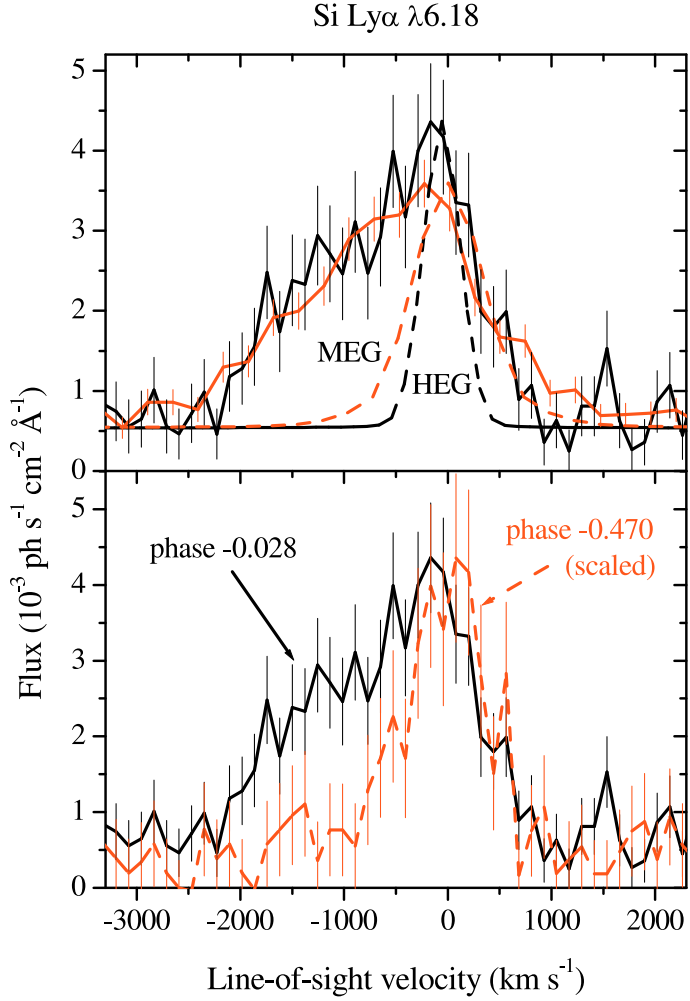}
  \bigskip
  \caption{Velocity profiles of the Si$^{+13}$ Ly$\alpha$ unresolved doublet
(6.180 and 6.186 \AA). {\it Upper panel} shows consistent HEG
(higher spectral resolution) and MEG (smoother) data from Obs.
3745 ($\phi$=~--0.028). The profile is clearly broadened up to
2000~\kms, much beyond the instrument line spread functions
(dashed lines). {\it Bottom panel} shows the HEG profile of Obs.
632 ($\phi$=~--0.470, dashed line) and Obs. 3745
($\phi$=~--0.028). The scaled up line from the early phase is
rather symmetrical and much narrower, showing no line emission
beyond 700~\kms.}
  \label{f1}
\end{figure}

%

\clearpage

\begin{figure}
\centerline{\includegraphics[width=10cm,angle=0]{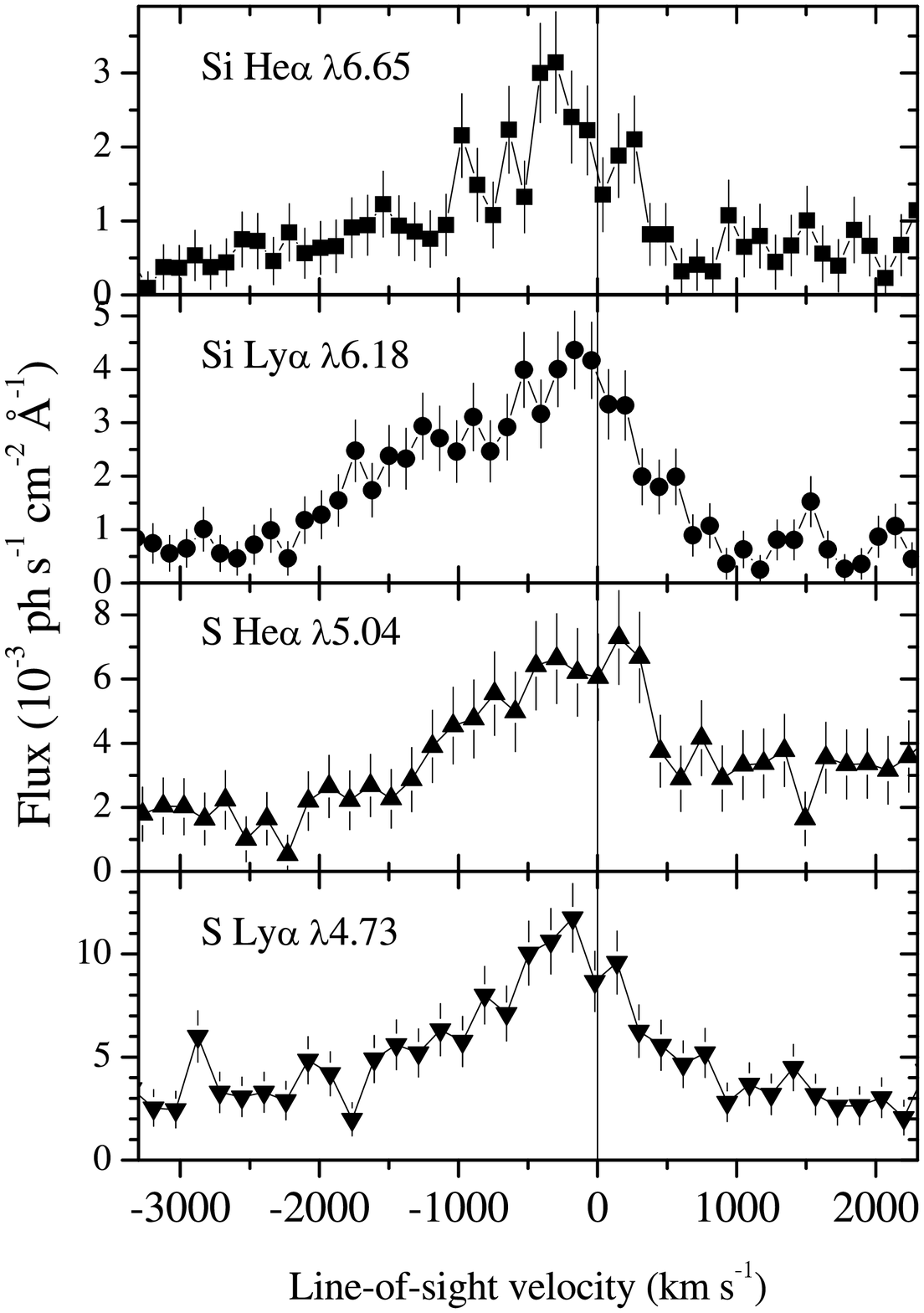}}

  \caption{Profiles of four bright spectral lines from Obs.~3745 (assumed phase --0.028).
  The consistent asymmetric profiles with blueshifts up to $\sim\ -2,000$~\kms\
  are clearly seen.}
   \label{f2}
\end{figure}

\clearpage


\begin{figure}
  \centering
  \includegraphics[width=10.0cm, angle=-90]{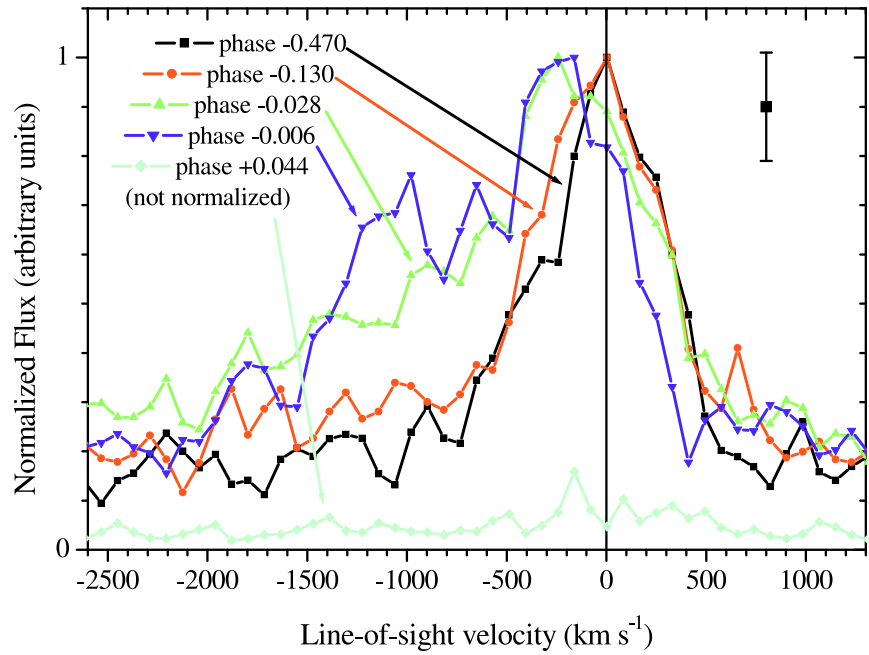}
  \bigskip
  \caption{Mean, normalized, velocity profiles constructed from nine different
  spectral lines for the five observations of \etc.
  Only the profile of Obs. 3747 (assumed phase +0.044) during which \etc\ was strongly
  absorbed is not normalized and plotted at its correct scale with respect to Obs. 632
 (assumed phase --0.470). Typical 1$\sigma$ errors on each data point are 10 -- 15~\%
 (top right hand side).
}
  \label{f3}
\end{figure}

\end{document}